\documentclass[preprint]{vgtc}               

\graphicspath{{figures/}{pictures/}{images/}{./}} 

\usepackage{times}                     

\usepackage{tabu}                      
\usepackage{booktabs}                  
\usepackage{lipsum}                    
\usepackage{mwe}                       

\usepackage{mathptmx}                  

\usepackage{calc}
\usepackage{caption}
\usepackage{subcaption}

\onlineid{0}

\vgtccategory{Research}

\vgtcinsertpkg

\preprinttext{This is the author's version, to appear in the IEEE VR 2025 conference.}

\title{Investigating the Influence of Playback Interactivity during Guided Tours for Asynchronous Collaboration in Virtual Reality}

\author{Alexander Giovannelli $^1$ \thanks{e-mail: agiovannelli@vt.edu} %
\and Leonardo Pavanatto $^1$ 
\and Shakiba Davari $^1$ 
\and Haichao Miao $^2$ 
\and Vuthea Chheang $^2$ 
\and Brian Giera $^2$ 
\and Timo Bremer $^2$ 
\and Doug A. Bowman $^1$} 
\affiliation{
    \scriptsize $^1$ Center for Human-Computer Interaction, Virginia Tech, USA \\ \scriptsize $^2$ Lawrence Livermore National Laboratory, USA %
}

\teaser{
  \centering
  \includegraphics[width=\linewidth]{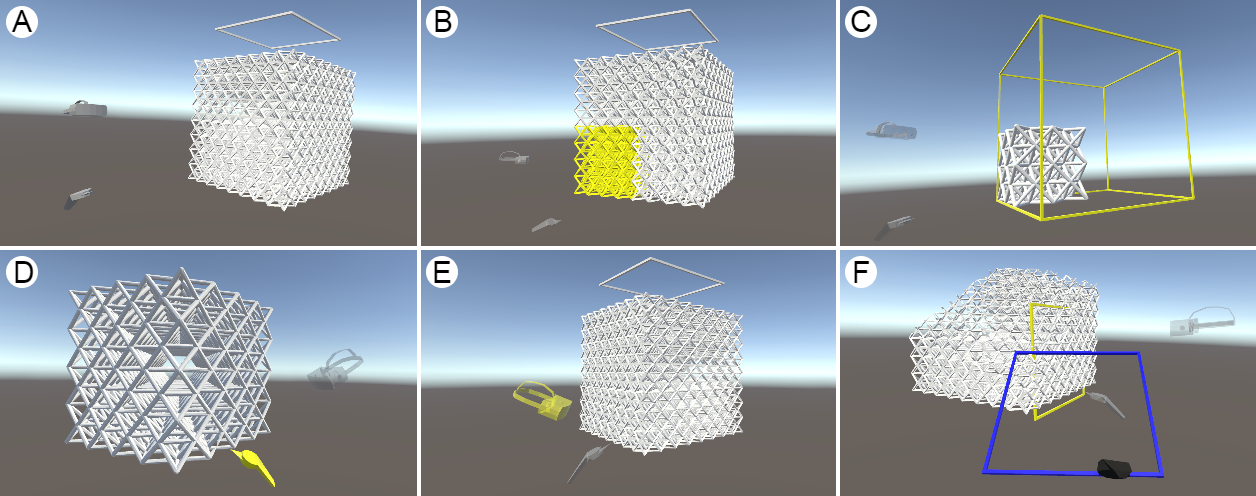}
  \caption{View of the inspection environment from the participant's field of view with corresponding yellow pause interaction indicators. The guide is the gray avatar, whose opacity is reduced during a pause requiring a given playback interaction. A: Standard playback. B: \textit{Navigation Down} interaction. C: \textit{Navigation Up} interaction. D: \textit{Controller Point} interaction. E: \textit{WYSIWIS} interaction. F: \textit{Plane Dock} interaction.}
  \label{fig:teaser}
}

\abstract{
    Collaborative virtual environments allow workers to contribute to team projects across space and time.
    While much research has closely examined the problem of working in different spaces at the same time, few have investigated the best practices for collaborating in those spaces at different times aside from textual and auditory annotations.
    We designed a system that allows experts to record a tour inside a virtual inspection space, preserving knowledge and providing later observers with insights through a 3D playback of the expert's inspection.
    We also created several interactions to ensure that observers are tracking the tour and remaining engaged.
    We conducted a user study to evaluate the influence of these interactions on an observing user's information recall and user experience.
    Findings indicate that independent viewpoint control during a tour enhances the user experience compared to fully passive playback and that additional interactivity can improve auditory and spatial recall of key information conveyed during the tour.
} 

\keywords{Virtual reality, computer-supported cooperative work, collaborative virtual environments, asynchronous collaboration, guided tours.}

\begin{document}


\maketitle

\section{Introduction} 
As Virtual Reality (VR) head-worn displays (HWDs) become more commercially viable in terms of cost and quality, designers and developers must consider the technology's most impactful application spaces.
An area of interest over the years has been collaboration within immersive environments, where a plethora of new cooperative capabilities can be realized \cite{Ghamandi_CollaborationTaxonomy2023}.
Coined as Collaborative Virtual Environments (CVEs), these digital landscapes allow multiple users to communicate with one another and interact with digital content through various interaction methods \cite{Churchill_CollabVEs1998, Steed_collaboration2015, Thomas_HybridFw2023}.
As a result of the geographical distribution of work becoming more common \cite{Tijssen_ResearchCollab2012} in recent years, telepresence tools such as Zoom\footnote{\url{https://zoom.us/}} and WebEx\footnote{\url{https://www.webex.com/}} have been used to facilitate remote collaboration.
Although these desktop-based collaboration applications help facilitate basic screen sharing and whiteboarding, they forfeit the benefit of embodied interaction, representation of information, and collaborators that CVEs provide for cooperative work \cite{Mayer_CollaborativeWorkImmersive2023, Giovannelli_Gestures2023}.

While a large amount of literature has focused on addressing the challenges of CVEs in the realm of synchronous collaboration \cite{Ens_RevisitingCollabMR2019}, using CVEs for asynchronous collaboration has been identified as an area of limited exploration \cite{Irlitti_ChallengesAsync2016, Pidel_CollabSystematic2020}.
CVEs have the potential to preserve increments of work \cite{Zhang_VRGit2023}, allow the annotation of knowledge and reasoning from individual contributions \cite{Marques_RemoteAsyncCollab2021}, draw collaborator attention and awareness to these annotations \cite{Tam_Framework2006}, and enable ad-hoc communicative exchange and relaying of pertinent information to drive continuous work across all hours \cite{Marques_AsyncNotifications2022}. 
These innate features of CVEs make them rich for further research investigation in their applicability for asynchronous collaboration. 

We propose extending asynchronous collaboration research in VR via guided tours.
Guided tours create an observer-guide relationship within immersive environments to allow narrative relaying of information across both space and time \cite{johansen:1989}.
Within a CVE, this would empower the preservation of expert knowledge regarding changes inside the environment made during a working session or the creation of overview information for novice observing users to follow later.
We motivate our work in the additive manufacturing (AM) specialty, falling within previously identified areas of potential asynchronous collaboration in engineering \cite{Mayer_CollaborativeWorkImmersive2023}.
During AM, inspection processes occur regularly, requiring experts across design, engineering, and manufacturing areas to provide feedback to ensure the quality of the produced part \cite{Chheang_AMInspect2024}.
However, whether from geographical distribution or scheduling constraints due to other responsibilities, these experts may not be able to inspect a given part jointly.
By using an inspection CVE, these experts can guide later observers through their inspection findings, preserving their analyses for continued collaboration asynchronously.

While in some respects viewing guided tours in a CVE is like a synchronous collaboration, the observer cannot directly communicate with the tour guide, nor can they influence the content of the tour.
This is analogous to the difference between attending an in-person lecture versus viewing a prerecorded lecture video.
Thus, we expect there might be a higher likelihood for observers to lose engagement with the tour, or to fail to completely follow the tour's informational content.
We hypothesize that embedding opportunities for the observer to interact with the playback or the environment during the tour will mitigate these effects and will lead to better information retention and user experience.

This paper details our design of a guided tour system with playback interactivity and a subsequent user study investigating its use for asynchronous collaboration in the context of AM.
We detail our findings regarding the use of interaction as a means to view and control the playback of the guided tour and its influence on the user's experience and retention of information covered in the tour's narrative.
We synthesize the findings with participant feedback and open-ended discussion, supporting the use of interactivity during playback when conducting guided tours instead of passive viewership. 
We also document the perceived benefits of interaction for playback control.
Contributions of this work include (1) the design of interactive methods to maintain observer engagement during guided tours in a CVE; (2) quantitative measurement of four varying increments of interactivity on user recall and user experience using a guided tour playback system; and (3) qualitative feedback regarding the design of interactive behaviors in a guided tour playback system.

\section{Related Work}

\subsection{Guided Tours}
Using ``The River Analogy,'' Gaylean proposed author-controlled navigation of virtual environments (VEs) to empower the structuring of experiences for informative presentation \cite{Galyean_GuidedNavigation1995}.
The analogy itself consisted of navigation akin to riding a boat on a river, where the users on the boat have minor control of direction, but ultimately the author controls the current by which the users move downstream and are shown the surrounding landscape.
While this did present a novel means of presenting information, the extent of control for users was reduced to allow guiding with no interruption.

Pausch et al. highlighted that such an approach would detract from the overall experience, where users would be unimpressed by the technology being used for its own sake; they care more about what there is that they can do in the virtual world \cite{Pausch_Aladdin1996}.
Through developing a VR system and evaluating it at a theme park, they found that most users preferred complete control of navigation, navigating to their own pointed places of interest in the experience.
This same user-controlled approach was applied in a PC-based experience for physicians to tour a patient's colon, allowing them to perform non-invasive colonoscopy procedures \cite{Hong_VirtualVoyage1997}.
The usage of guided navigation was found to empower computer-illiterate surgeons to successfully inspect areas of interest inside the colon by instead using mouse peripherals to view locations of interest.

Best further extended the concept of guided tours in the context of museums, proposing digital guides to deliver personalized tours for users \cite{Best_MuseumTours2012}.
These digital guides would act as a pathfinder, leading users around a unique site, and a mentor, providing key information to users about that site.
However, Best's insights were gleaned from her own observations of recorded, real-world museum tours, where Tsiropoulou et al. proposed a human-in-the-loop approach \cite{Tsiropoulou_ExpMuseumTour2017}.
They highlight the importance of capturing the reported mental, physical, and spatial impacts of visitors' overall quality of experience obtained from visiting individual exhibits.\\

\noindent From this existing literature, work regarding guided tours has primarily focused on controlling the presentation of data or knowledge by managing the extent of an individual's ability to navigate and interact inside virtual and physical space. 
While the potential use of digital guides to steer observers to points of interest and provide key information is addressed, no resulting evaluation or use is investigated.
We explore this digital guide concept in our work as a means to not only control the flow of information shared but also to measure the impact on recall and experience.

\subsection{Annotations}
Early usage of VEs focused on their capability to not only serve as simulator systems but to empower the marking of a VE to preserve information for future review by a peer or collaborator using annotations.
Verlinden et al. described such a system, noting the possibilities of using annotations to link temporal features of a VE \cite{Verlinden_VirtualAnnotation1993}.
They noted the importance of documenting findings associated with various data objects such that ``notes are spatially indexed'' and that, within the simulation, those annotations should be ``natural'' and create new ways to communicate with others.
A similar system was introduced by Harmon et al., adding the aforementioned annotation association capabilities by supporting ``object annotation'' and ``view annotation'' \cite{Harmon_AnnotationSys1996}. 

Frécon \& Nöu leveraged these annotation methods explicitly for collaboration by creating a VE modeled after a meeting room and providing a set of virtual tools allowing users to store text artifacts (e.g., notebooks and reports) and further annotate those artifacts \cite{Frecon_BuildingDistVEs1998}.
Since these early works, annotations have been investigated thoroughly, identifying characteristics such as placement methods \cite{Bowman_IRVE2003, Wither_OutdoorAR2009, Pick_PosAnnotation2010} and media types of text \cite{Lisle_ISTSense2021, Tahmid_ColtCollab2023}, speech \cite{Verlinden_VirtualAnnotation1993, Ribarsky_VRSpeech1994}, images \cite{Tahmid_ISTClusters2022}, videos \cite{Hansen_UbiAnnotation2006}, sketches \cite{Guerreiro_BeyondPostIt2014, Clergeaud_AnnotationSys2017}, and even world-views \cite{Bell_ViewMgmt2001}.\\

\noindent Although these researched systems and types of annotations provide features that theoretically support collaboration, they do not explicitly investigate the implications of use for asynchronous collaboration.
In our work, we investigate using guiding avatars as a means to annotate information in the ``natural'' way.

\subsection{Asynchronous Collaboration}
Asynchronous collaboration's flexibility to allow contribution of work regardless of time is its greatest strength compared to its synchronous counterpart \cite{Irlitti_InteractAsync2013}.
However, while research has largely solved the authoring of annotations for asynchronous collaboration, the complexity of reviewing these annotations at a later time and the resulting understanding from a review has been seldom investigated \cite{Irlitti_ChallengesAsync2016}.
Although identified as a topic of limited research interest in VR \cite{Pidel_CollabSystematic2020}, asynchronous collaboration has seen many new contributions in recent years.

Wang et al. developed a communication tool in VR for 3D asynchronous collaborative design, comparing it with desktop-audio and desktop-based CVEs \cite{Wang_VRReplay2019}.
The system recorded and replayed avatars as they placed and moved furniture within a room, with participants preferring the use of VR in dimensions of communication clarity, perception of partners, performance satisfaction, and outcome satisfaction.
Chow et al. similarly contributed to the area of interior design, further analyzing the social behavior in asynchronous collaboration \cite{Chow_ChallengesAsyncCollabVR2019}.
They proposed design recommendations from their evaluation: providing rich navigation cues, activity highlighting, and animation changes at transitions in the CVE.

Outside the interior design and decorating use case, Marques et al. examined the use of notification sharing via spatial annotations in maintenance scenarios with augmented reality (AR) \cite{Marques_RemoteAsyncCollab2021, Marques_AsyncNotifications2022}.
Their system allowed onsite technicians to use a handheld device to annotate problems for remote experts to instruct them later on repair methods.
García-Pereira created a similar system, focusing on the quality of handheld devices used to create spatial annotations and to preserve position in search and find tasks for later users \cite{Garcia_CrossDeviceARAnnotations2021}.
Lee et al. also explored this concept of using AR spatial annotations to provide instructional videos on everyday tasks \cite{Lee_EnhancingInstructionVideos2020}.
Several works have supplemented these works by analyzing methods to assist in the authorship of spatial annotations, with Wang et al. exploring the use of freehand behaviors for context-aware creation and responsiveness of AR virtual contents \cite{Wang_CAPturAR2020, Wang_GesturAR2021}.

Expanding to asynchronous collaboration across the reality-virtuality continuum \cite{Milgram_RealityVirtuality1995}, Fender et al. created and analyzed a system that preserved co-located or remote interruptions from an observer of an occupied VR user \cite{Fender_CausalityAsync2022}.
A similar system was produced by Cho et al. for tracking and visualizing events in a mixed reality user's physical space to provide enhancements to memory awareness of these background occurrences \cite{Cho_RealityReplay2023}.

In the manufacturing space, Mayer et al. developed and evaluated a system that had a presenting avatar perform an assembly process \cite{Mayer_AsyncManualWork2022}.
Participants could follow this collaborative avatar as opposed to learning from a virtual manual to complete their own assembly.
They further noted the potential of using asynchronous collaboration systems in the digitalization of artifacts to empower remote work in business, education, and engineering sectors \cite{Mayer_CollaborativeWorkImmersive2023}.\\

\noindent While these works have examined asynchronous collaboration using VR technologies, they have focused on social outcomes, question-and-answer behaviors, and the performing of instructions with explicit playback controls.
Our work builds on this by evaluating the use of interactivity as a means to maintain engagement and understanding during an asynchronous inspection process.

\begin{table*}[]
\resizebox{\textwidth}{!}{%
\begin{tabular}{|l|l|r|}
\hline
\multicolumn{1}{|c|}{\textbf{Test A}}                                            & \multicolumn{1}{c|}{\textbf{Test B}}                           & \multicolumn{1}{c|}{\textbf{Question Type}} \\ \hline
What is the potential cause of bent and broken strut defects?                    & How many unit cells exist in the second level of the model?    & Verbal                                      \\ \hline
Place the red sphere at the location of the second defect discussed in the tour. & Place the red sphere at the location of the Bent strut defect. & Spatial                                     \\ \hline
What is the second defect referred to as?                                        & Which defect is the most severe?                               & Verbal                                      \\ \hline
Place the red sphere at the location of the Broken strut defect.                 & Place the red sphere at the location of the Thin strut defect. & Spatial                                     \\ \hline
How does a 3D printer create a thin strut defect?                                & What are the individual unit cells referred to as?             & Verbal                                      \\ \hline
\end{tabular}%
}
\setlength{\belowcaptionskip}{-5pt}
\vspace*{-5pt}
\caption{Test questions in Test A and Test B.}
\label{tbl:tests}
\end{table*}

\section{Guided Tour Playback Interaction Design} \label{sec:gtdesign}
We designed a system for a guided tour where a mentoring guide explained the inspection process of a 3D-printed object. 
The object within the VE was generated using open-source computed tomography data from AM octet lattice structures\footnote{\url{https://data-science.llnl.gov/open-data-initiative}} and the narrative spoken by the guide from the literature regarding these structures \cite{Klacanksy_InspectionAM2022}.
During the tour, the guide's voice was heard explaining the inspection verbally, and the guide's avatar (i.e., head and hands as shown in Fig. \ref{fig:teaser}A) was seen moving around the object, looking or pointing at various parts of the object, cutting into the object with a cutting plane, and navigating through different levels of scale of the object.

Our goal was to foster the observing user's engagement in the guided tour during the inspection process.
To achieve this goal, we designed interactions to make the user mimic the guide's actions during the tour.
To integrate these interactions into the tour, the playback would pause whenever the guide performed any of the following: navigation across scales, usage of tools, and view of regions or cells of interest on the 3D-printed object.
The playback would resume once the user performed the corresponding interaction based on the guide's actions.
The guide is responsible for specifying the interactions they want the observing user to mimic by using explicit controller input in a separate recording mode available to designated guide users.

For these playback interactions, visual and auditory feedback indicated to the user that an interaction was required.
The audio cue consisted of a unique sound per interaction and was played at the user's position.
The visual indicator included an object in the environment spawning with a yellow material, highlighting the required interaction for continued playback.
Additionally, the guide's avatar turned from gray to transparent gray while the playback was paused, returning to gray once the required interaction was performed and playback resumed.
The following details the mimicking interactions in generalized groups.\\

\noindent\textbf{Multiscale Navigation}
During the inspection, the guide traverses across scales of the 3D-printed object, effectively zooming in and out of specified sections.
This provides a closer, detailed view of a smaller group of cells comprising the print versus a wider, less detailed view of many cells.
We divide multiscale navigation capability across two interactions: \textit{Navigate Down} and \textit{Navigate Up}.

\textbf{\textit{Navigate Down:}}
When the guide navigates down, their avatar shrinks in scale by a multiplier of 0.5.
The octant they selected for the navigation down then becomes yellow, as shown in Fig. \ref{fig:teaser}B.
The user must then perform the same narrowing navigation by aiming at the highlighted octant and pressing the controller’s trigger to continue the tour playback.
This reduces the player's scale by the 0.5 multiplier and removes the cells outside of the selected octant to disappear from the environment.
The number of unit cells displayed when navigating down in scale is 512 to 64 and 64 to 8.

\textbf{\textit{Navigate Up:}}
When the guide navigates up, their avatar grows in scale by a multiplier of 2.
The position in the print that they have navigated up toward is represented by a yellow box, as shown in Fig. \ref{fig:teaser}C.
The user must then perform the same upward navigation by pressing the primary button on the controller.
This grows the player's scale by the multiplier of 2, populates the yellow grid with the cells comprising it, and resumes the tour playback.
The number of unit cells displayed when navigating up in scale is 8 to 64 and 64 to 512.\\

\noindent\textbf{Perspective Sharing}
During the inspection, the guide points toward and looks at regions, features, or specific cells comprising the 3D-printed object, calling for the user to pay attention to them.
This causes the user to identify these features or specific cells by visually directing their attention toward them before a verbal explanation is given describing their significance.
We divide the visuospatial relation of described features or cells across three interactions: \textit{Controller Point}, \textit{WYSIWIS}, and \textit{Plane Dock}.

\textbf{\textit{Controller Point:}}
Whenever the guide pointed to a region or cell of interest that required the user's attention, the guide's controller would stop before that region or cell and turn yellow, as shown in Fig. \ref{fig:teaser}D.
The user must then observe the point of interest specified by the guide's controller.
They then aimed at the guide's controller, pressing their own controller's trigger to acknowledge the point of interest and resume tour playback.

\textbf{\textit{WYSIWIS:}}
When the guide wanted the user to observe the 3D-printed object from a specific point of view, the guide's head would stop at that viewpoint and turn a transparent yellow, as shown in Fig. \ref{fig:teaser}E.
The user then needed to position their head within the guide avatar's head to see from their perspective, aligning with its approximate rotation for 1.5 seconds.
After completing this interaction, known as ``What You See Is What I See'', the tour playback resumed \cite{Stefik_Wysiwis1987}.

\textbf{\textit{Plane Dock:}}
A cutting plane in the environment allowed the guide to cut through the 3D-printed object, revealing its inner structure, removing occluding cells, and providing a desired view of the remaining cells.
To have the user understand the positioning of the plane and the resulting desired view of the cells, the guide spawned a yellow dock at the position and rotation of a required cut, as shown in Fig. \ref{fig:teaser}F.
The user then needed to grab the cutting plane by hovering over it with their controller and pressing the trigger on the controller, then placing it within the approximate position and rotation of the yellow dock.
After releasing the cutting plane in the dock, the tour playback resumed.

\section{Experiment}
Our experiment described in this section was designed to investigate the impact of interactivity during guided tour playback for asynchronous collaboration. 
More specifically, we aimed to better understand how the extent of interaction afforded to an observer influences them following a tour guide presenting information in a VE.
For this reason, we designed an experiment to address the following research questions:
\begin{enumerate}
    \setlength\itemsep{0em}
    \item[\textbf{RQ1}] How does the extent of interaction during an immersive guided tour impact an observing user's recall?
    \item[\textbf{RQ2}] How does the extent of interaction during an immersive guided tour impact an observing user's experience?
\end{enumerate}

\subsection{Task}
Participants were instructed to act as collaborative observers on a tour where the guide explained the inspection process of a 3D-printed object. 
During the tour, the guide explained different defects and their root causes, showing examples of the defects in the 3D-printed object.
The participants were tasked with following the guide's verbal and visuospatial inspection and were informed they would be tested on both the verbal and spatial content of the tour.

\subsection{Measures} \label{sec:measures}
Various quantitative and qualitative measures were collected during the study.
The Simulator Sickness Questionnaire (SSQ) was administered to measure any changes to the participant's symptoms in oculomotor, disorientation, and nausea \cite{Kennedy_SSQ1993}.
To assess the workload of the guided tour across the conditions, the NASA Task Load Index (NASA-TLX) was utilized \cite{Hart_NASATLX2006}.
The User Engagement Scale Short Form (UES-SF) was used to measure the participant's engagement, specifically, their focused attention, perceived usability, aesthetic appeal, and reward factors from following along with the guided tour \cite{Obrien_UES2018}.
Two custom tests were administered to gauge participants' knowledge of the verbal and spatial information covered during the guided tour.
We detail the questions for each test in Table \ref{tbl:tests}.
Finally, an audio-recorded interview was performed after the tour to acquire open-ended feedback on the participant's experience of their tour condition.
During the audio-recorded interview, we prompted participants for a 1-10 rating as to whether they would like to experience their tour condition again with a topic they were passionate about instead.
The other questions used for the interview are listed in Table \ref{tbl:interview}.

\begin{table*}[]
\resizebox{\textwidth}{!}{%
\begin{tabular}{|l|}
\hline
\multicolumn{1}{|c|}{\textbf{Interview Questions}}                                                                                                                                                                                                                                                                                                \\ \hline
\begin{tabular}[c]{@{}l@{}}Suppose that someone wanted to teach you about a topic you are passionate about using a guided tour system similar to the one you just experienced. \\ On a scale from 1-10, with 1 being 'not at all' and 10 being 'very much', how likely would you be to take another guided tour in this format? Why?\end{tabular} \\ \hline
Considering all the ways the tour tried to draw your attention to areas of interest, which interactions by the guide were helpful to you following along? Why?                                                                                                                                                                                    \\ \hline
Were there any interactions you would like to be able to do inside the tour that you did not experience today? If so, describe them.                                                                                                                                                                                                              \\ \hline
Did you encounter any issues when following the tour? If so, describe them.                                                                                                                                                                                                                                                                       \\ \hline
\end{tabular}%
}
\setlength{\belowcaptionskip}{-5pt}
\vspace*{-5pt}
\caption{Interview questions asked during the post-study.}
\label{tbl:interview}
\end{table*}

\subsection{Conditions} \label{sec:conditions}
Four conditions were created encapsulating incremental combinations of our playback interactions (see Sec. \ref{sec:gtdesign}). 
We detail these conditions in order of least to most interactivity during playback below.

The \textsc{Passive} condition did not include any playback interactions for the user to perform.
The tour showed the first-person point of view of the guide in a virtual display positioned 1m away from the user in an empty VE.
This condition was intended to have users strictly follow the inspection from the guide's perspective and to distinguish the implications of no viewpoint control in the VE compared to the other conditions.

The \textsc{Rails} condition also did not include any playback interactions for the user to perform.
However, the user had full control over their viewpoint within the VE, meaning they needed to maneuver themselves to follow the guide during the tour.
For example, whenever the guide navigated upward or downward, the system automatically scaled the user, requiring them only to physically adjust their orientation to effectively follow the tour.

The \textsc{Navigation} condition consisted of the \textit{Multiscale Navigation} playback interactions.
Playback of the tour would pause until \textit{Navigate Down} and \textit{Navigate Up} interactions were performed by the user to follow those of the guide.

Finally, the \textsc{Interactive} condition used the \textit{Multiscale Navigation} and \textit{Perspective Sharing} playback interactions.
Playback of the tour would pause until \textit{Navigate Down} and \textit{Navigate Up} interactions were performed by the user to mimic the same actions of the guide.
Additionally, tour playback would pause whenever the guide specified a point of interest requiring the user's attention using the \textit{Controller Point}, \textit{WYSIWIS}, and \textit{Plane Dock} interactions.

\subsection{Hypotheses}
We created hypotheses for each of the research questions before conducting our experiment.
Regarding the influence of interaction on participant recall (\textbf{RQ1}), we hypothesized:
\begin{enumerate}
\setlength\itemsep{0em}
\item[\textit{H1.1}] The \textsc{Interactive} condition will have a greater recollection of verbal information than all other conditions. 
\item[\textit{H1.2}] The \textsc{Interactive} condition will have a greater recollection of spatial information than all other conditions.
\end{enumerate}

\noindent From a social psychological perspective of the tourist-guide relationship, we expect internal mental retention of information accompanying interaction within a given spatial environment \cite{Argyle_Social1981, Pearce_TouristGuideInteraction1984, Thagard_CogSci2023}.\\

\noindent In regards to the influence of interaction on the participant's experience (\textbf{RQ2}), our hypotheses are as follows:
\begin{enumerate}
\setlength\itemsep{0em}
\item[\textit{H2.1}] The \textsc{Interactive} and \textsc{Navigation} conditions will have a higher reported engagement than the non-interactive conditions.
\item[\textit{H2.2}] The \textsc{Passive} condition will be reported as having a higher task workload than all other conditions.
\item[\textit{H2.3}] The \textsc{Rails} and \textsc{Passive} conditions will be reported as having higher simulator sickness scores.
\end{enumerate}

\noindent From prior work targeting visualization and interaction within VR guided tours, we expect favorable reviews as a result of more interaction performed by a given participant \cite{Letellier_VizInteractionGuidedTours2019, Pausch_Aladdin1996}.

\subsection{Design}
The experiment used a between-subject design, where each participant observed a tour for only one of the conditions described in Section \ref{sec:conditions}.
Participants were randomly assigned to one of the four conditions, so each condition had 10 participants. 
The administration of tests described in Section \ref{sec:measures} was counterbalanced, with half of the participants completing Test 1 first followed by Test 2 and vice versa.
The independent variable was the extent of interaction afforded to the user during the tour.
Dependent variables were the participant's tour test scores, and subjective responses for the SSQ, UES-SF, NASA-TLX, and audio-recorded interview.

\subsection{Participants} \label{sec:participants}
A total of 40 participants were recruited for the study using university email lists and newsletters.
All were screened before participating, indicating they had normal or corrected vision, were comfortable standing for long periods, and were fluent in English.
Additionally, two Likert scales between 1-6 (i.e., 1 being ``not'' and 6 being ``very'')  were completed regarding the user's interest in and familiarity with 3D printing.
For the \textsc{Interactive} condition, four participants were female and six were male, with ages ranging between 19 to 30 years (\emph{M}=24.70, \emph{SD}=3.47).
Seven participants reported having experienced VR 10 or more times, one between 3-10 times, and two between 0-2 times.
Likert scale values reported a mean of 2.5 for familiarity and 3.4 for interest in 3D printing. 
In the \textsc{Navigation} condition, four participants were female and six were male, with ages ranging between 20 to 36 years (\emph{M}=28.10, \emph{SD}=5.47).
Five participants reported having experienced VR 10 or more times, four between 3-10 times, and one between 0-2 times.
Likert scale values reported a mean of 2.6 for familiarity and 3.9 for interest in 3D printing. 
For the \textsc{Rails} condition, four participants were female and six were male, with ages ranging between 23 to 35 years (\emph{M}=28.00, \emph{SD}=3.97).
Nine participants reported having experienced VR 10 or more times and one between 3-10 times.
Likert scale values reported a mean of 3.4 for familiarity and 4.3 for interest in 3D printing. 
Finally, in the \textsc{Passive} condition, four participants were female and six were male, with ages ranging between 23 to 35 years (\emph{M}=27.60, \emph{SD}=4.38).
Six participants reported having experienced VR 10 or more times, two between 3-10 times, and two between 0-2 times.
Likert scale values reported a mean of 2.5 for familiarity and 4.2 for interest in 3D printing. 

\begin{figure*}[tb]
    \centering
    \includegraphics[width=\textwidth]{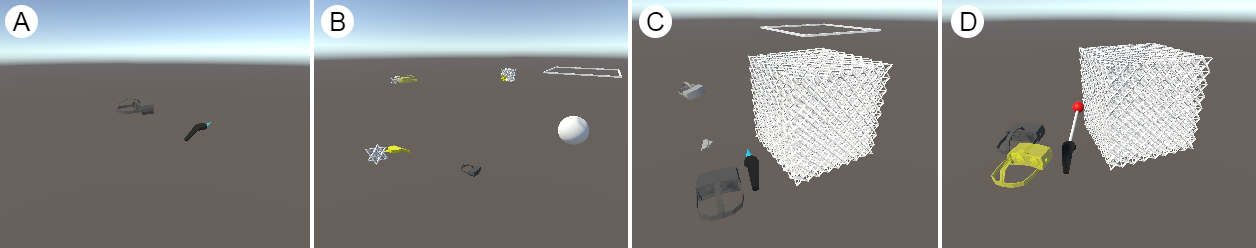}
    \caption{Exocentric views of the VEs used in the experiment. A: Setup VE showing the participant's avatar in its empty skybox. B: Training VE showing the props used to teach all playback interactions for the \textsc{Interactive} condition. C: Inspection VE showing the participant's avatar observing the guide's avatar and the 3D-printed object. D: Test VE showing the player avatar performing a spatial marking task using the `marking wand'.}
    \label{fig:ves}
    \vspace*{-15pt}
\end{figure*}

\subsection{Apparatus \& Environment} \label{sec:appAndVe}
Participants wore a Meta Quest 2 HWD during the study. 
It has a resolution of 1832x1920 per eye with a refresh rate of 90Hz and a 104-degree horizontal and 98-degree vertical Field of View (FoV) \footnote{\url{https://www.meta.com/quest/products/quest-2/}}.
A Meta Quest 2 Touch controller, held in the participant's specified dominant hand, was used to perform the interactions described in Section \ref{sec:conditions}.
The participant's movement within a VE was tracked using the Meta Quest Headset Tracking system\footnote{\url{https://www.meta.com/help/quest/articles/headsets-and-accessories/using-your-headset/}}.
The boundary was created in an unobstructed area of approximately 2.5m x 2.5m and calibrated using floor markers in the physical environment before conducting each study.

Four VEs were used during the study: Setup, Training, Inspection, and Test.
The Setup VE was an empty skybox environment and was strictly used to perform height calibration of the participant and facilitate swapping between the other VEs by the investigator.
The Training VE was populated with props based on the interaction condition assigned to the participant.
The Inspection VE had a cuboid lattice model with 1m x 1m x 1m dimensions in its center.
This same environment was used for the Test VE, however, controls were added for the investigator to administer test questions.
Additionally, a transparent yellow headset was shown in the Test VE, marking where the participant started the inspection tour.
These VEs are shown in Fig. \ref{fig:ves}.

\subsection{Procedure} \label{sec:procedure}
Each participant completed three phases as part of the experiment: pre-study, study, and post-study.
In the pre-study, the participant read and signed an informed consent document about the study and completed an SSQ.
Then the participant was introduced to the HWD and had their interpupiliary distance measured so that the lens positions could be set accordingly in the HWD.
Finally, the participant was asked their dominant hand and shown the corresponding controller and the buttons they would need to use during the post-study and/or study according to their interaction condition.
Once they indicated they were comfortable with the controller and understood how to adjust and put on the HWD, the study phase started.

In the study phase, the participant had their height calibrated inside the Setup VE to a uniform height of 1m.
The investigator then placed the participant in the Training VE and started the training tour.
During the training tour, the participant followed a guide explaining the interactions to be expected during the inspection tour based on their study condition and was required to perform interactions to proceed in the tour accordingly (see Section \ref{sec:conditions}).
After completing it, the investigator answered any questions regarding the interactions, marking the end of the training tour.
The investigator then placed the participant in the Inspection VE.  
Before starting the inspection tour, the investigator gave a short description regarding the subject of the tour (i.e., the inspection of a 3D-printed object) and informed the participant that they would be tested regarding key locations shown and descriptions provided by the guide about the object from the tour.
The participant then followed the inspection tour, lasting approximately eight minutes.
The end of the inspection tour concluded the study phase and the participant then moved on to the final phase: the post-study.

In the post-study phase, the participant removed the HWD and completed another SSQ.
After approximately three minutes passed from removing the HWD, the investigator had the participant put the HWD back on and placed the participant in the Test VE.
The participant was then prompted to answer the questions from Table \ref{tbl:tests} for one of the tests, positioning themselves at the starting marker as described in Section \ref{sec:appAndVe} before being asked each question.
For verbal questions, the participant answered orally and the investigator wrote down their response for later grading.
For spatial questions, the participant used a wand with a red sphere at its tip attached to their controller to mark the approximate position of the prompted defect in the 3D-printed object at the highest scale, pulling the trigger once they believed the red sphere overlapped with that cell.
The distance between the red sphere and the target cell was automatically logged by the system.
Once they answered all the questions, the participant again removed the HWD and completed the UES-SF and NASA-TLX questionnaires.
After completing the questionnaires, the investigator conducted an audio-recorded interview with the participant regarding the guided tour they experienced.
Following the interview, the participant put the HWD back on and performed a buffer activity: the tutorial and demo song included in the Beat Saber - Demo\footnote{\url{https://www.meta.com/experiences/1758986534231171/}}.
This activity served to clear participants' working memory of tour information, having them remove their focus of attention on tour content and its required sensorimotor actions by performing the same actions in an unrelated experience \cite{Jonides_MindMemory2008}.
Once they completed the tutorial and demo, they took the remaining test in the Test VE.
In this way, we hoped to measure immediate recall with the first test and medium-term retention with the second test.
Finally, they completed a background questionnaire with demographic information as described in Section \ref{sec:participants}.
The entire study took approximately 60 minutes.

\section{Results}
We utilized various analyses to test our hypotheses using the data collected during our study. 
We first performed the Kruskal-Wallis tests to check for significance across dimensions for non-parametric Likert scale-based questionnaire responses.
If statistical significance was found, we then performed post-hoc pairwise comparisons using Mann-Whitney U tests with Holm-Bonferroni corrections.
We report this approach's statistical significance for the SSQ, UES-SF, NASA-TLX, and condition rating results.
Analysis methods for the test questions posed in Table \ref{tbl:tests} are explained in their respective subsections.

\subsection{Simulator Sickness}
We obtained the difference in reported simulator sickness by subtracting the pre-study SSQ from the post-study SSQ.
The \textit{Total} score was then calculated, as well as the separate sickness scores of \textit{Disorientation}, \textit{Nausea}, and \textit{Oculomotor} dimensions. 
We did not find any significant effects between the interaction conditions on simulator sickness increase across any of these dimensions from the Kruskal-Wallis tests performed.
We additionally analyzed the difference in pre- and post-study SSQ scores for each condition individually.
Mann-Whitney U tests revealed significant increases in simulator sickness scores of the \textsc{Passive} interaction condition for the \textit{Disorientation} ($\emph{p}<0.01$), \textit{Nausea} ($\emph{p}=0.03$), \textit{Oculomotor} ($\emph{p}<0.01$), and \textit{Total} ($\emph{p}<0.01$) dimensions.
The \textsc{Interactive} condition also had significant increases in simulator sickness scores for the \textit{Oculomotor} ($\emph{p}=0.03$) and \textit{Total} ($\emph{p}=0.04$) dimensions.
These significant pairs and the mean pre- and post-study sickness scores are shown in Fig. \ref{fig:ssq_plot}.

\subsection{Engagement}
\begin{figure*}[!t]
    \centering
    \begin{subfigure}[b]{\columnwidth}
        \centering
        \includegraphics[width=\columnwidth]{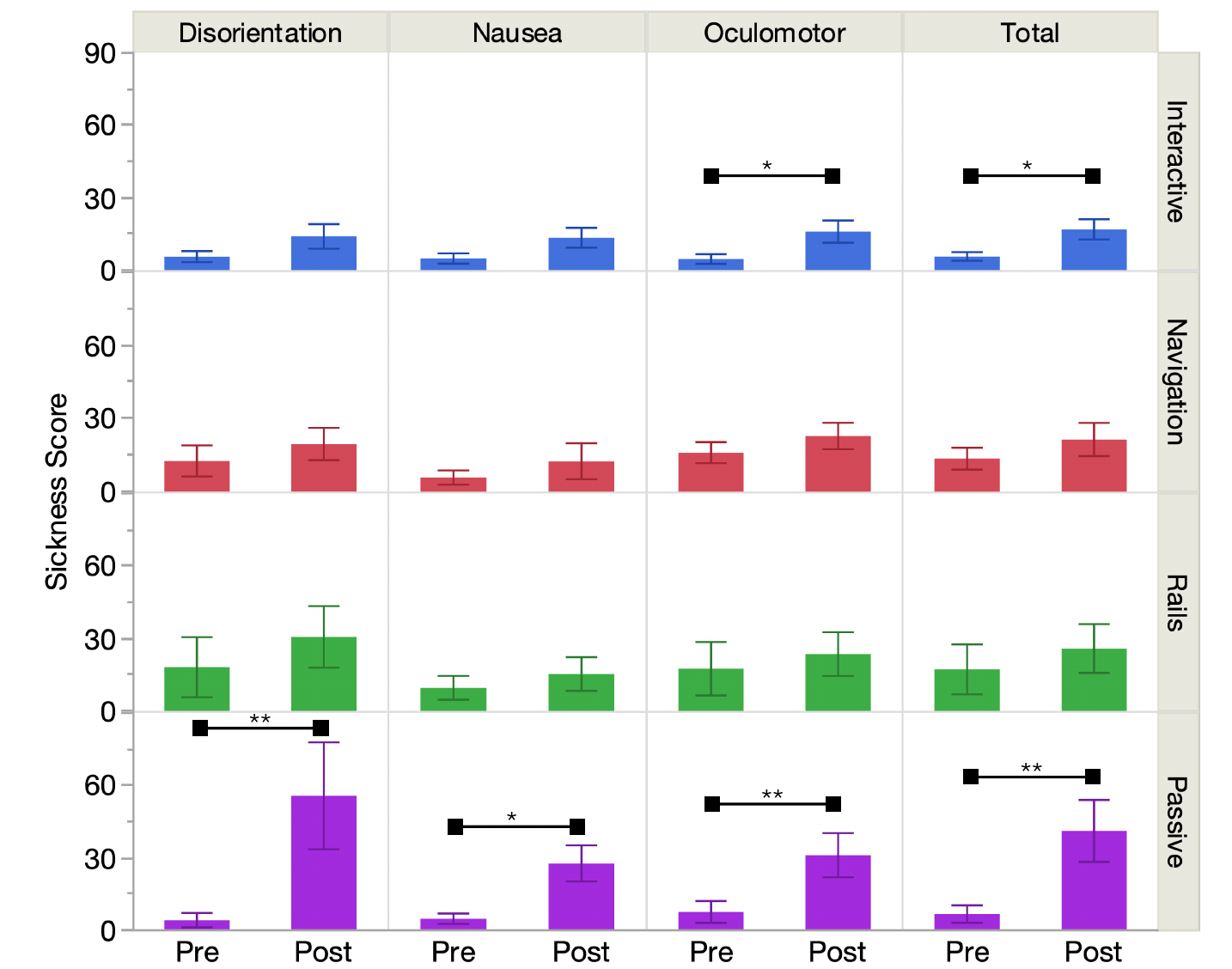}
        \caption{ }
        \label{fig:ssq_plot}
    \end{subfigure}
    \hspace{2.1em}
    \begin{subfigure}[b]{\columnwidth}
        \centering
        \includegraphics[width=\columnwidth]{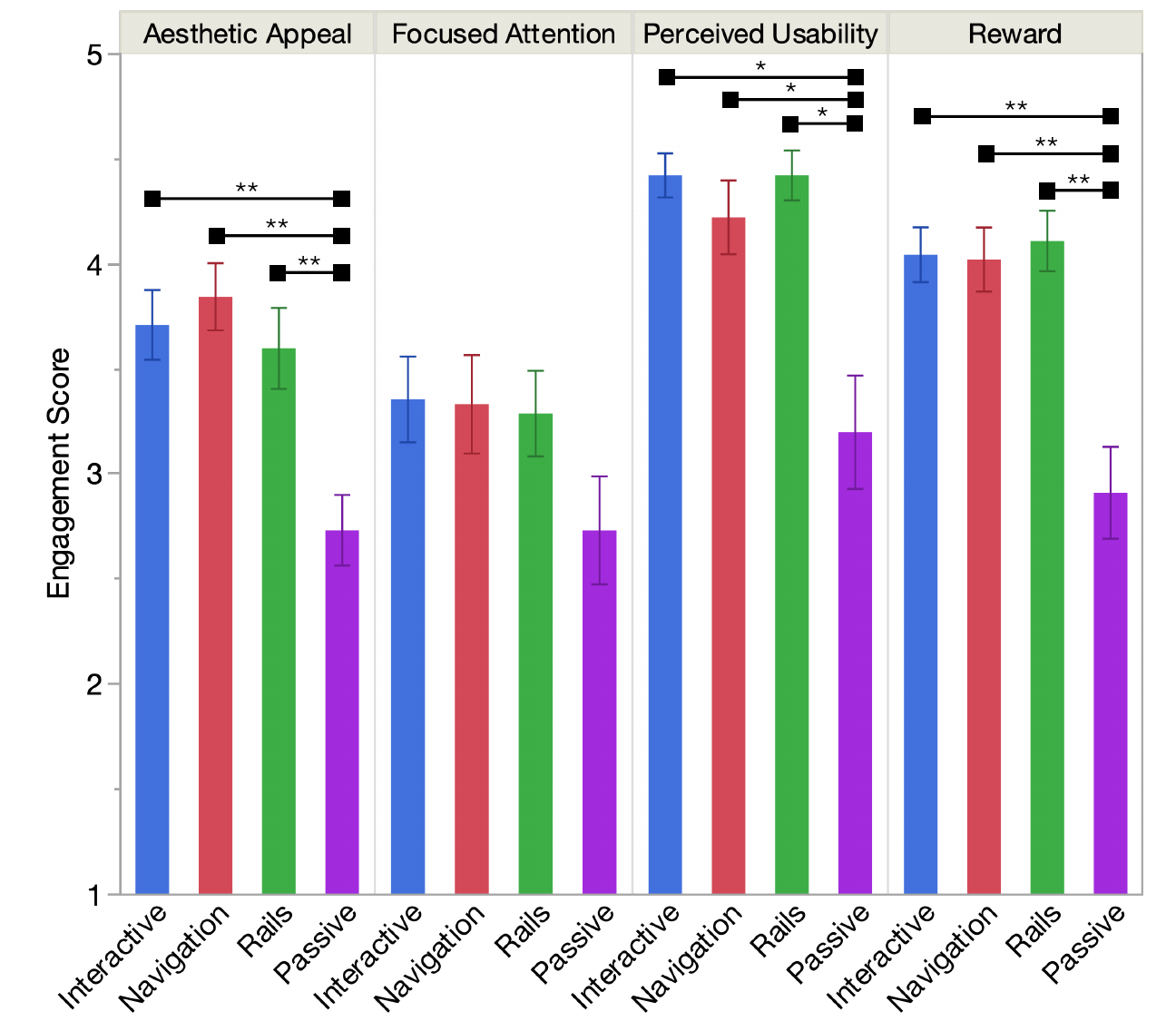}
        \caption{ }
        \label{fig:ues_plot}
    \end{subfigure}
    \begin{subfigure}[b]{\columnwidth}
        \centering
        \includegraphics[width=\columnwidth]{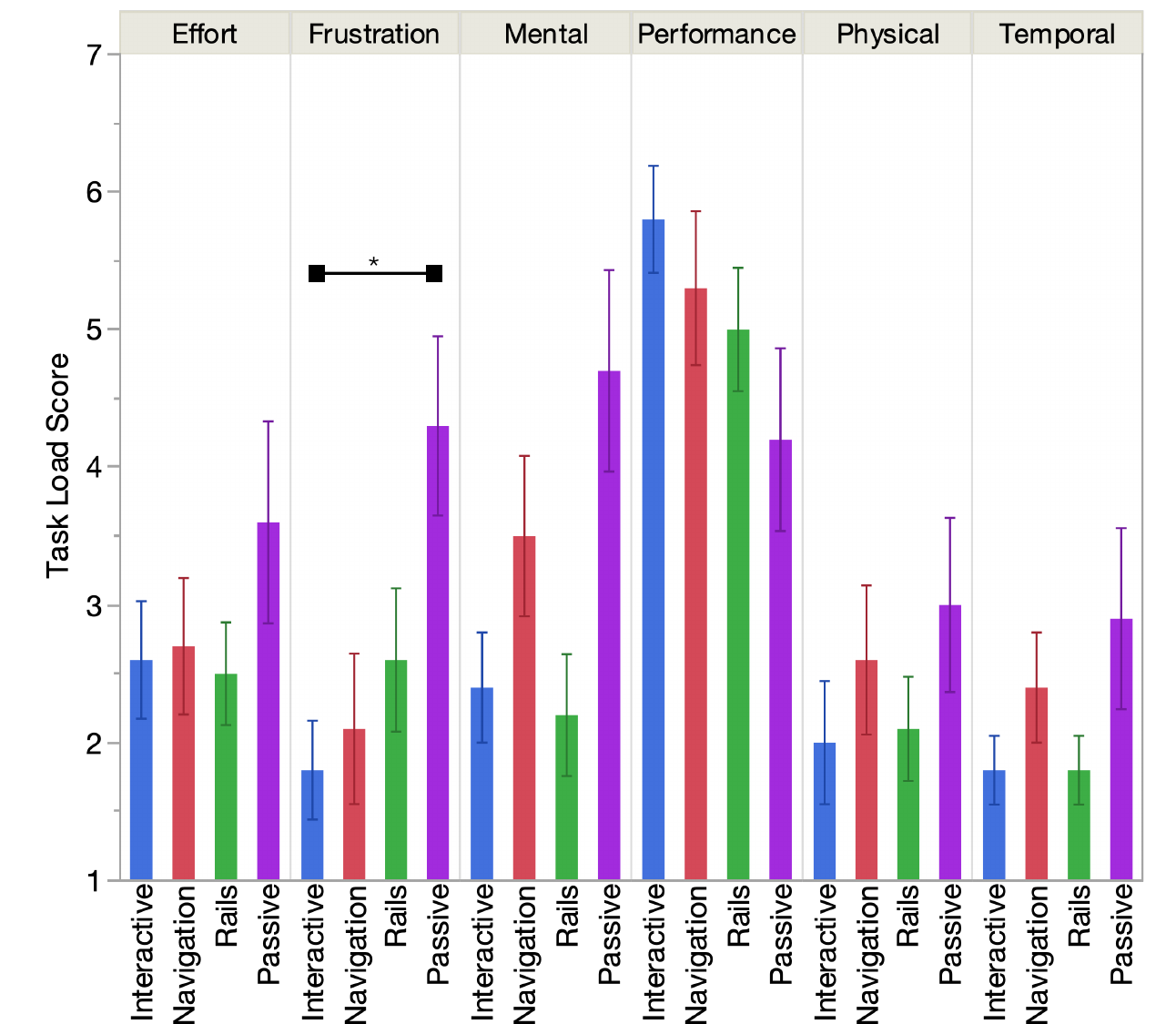}
        \caption{ }
        \label{fig:tlx_plot}
    \end{subfigure}
    \hspace{2.1em}
    \begin{subfigure}[b]{\columnwidth}
        \centering
        \includegraphics[width=\columnwidth]{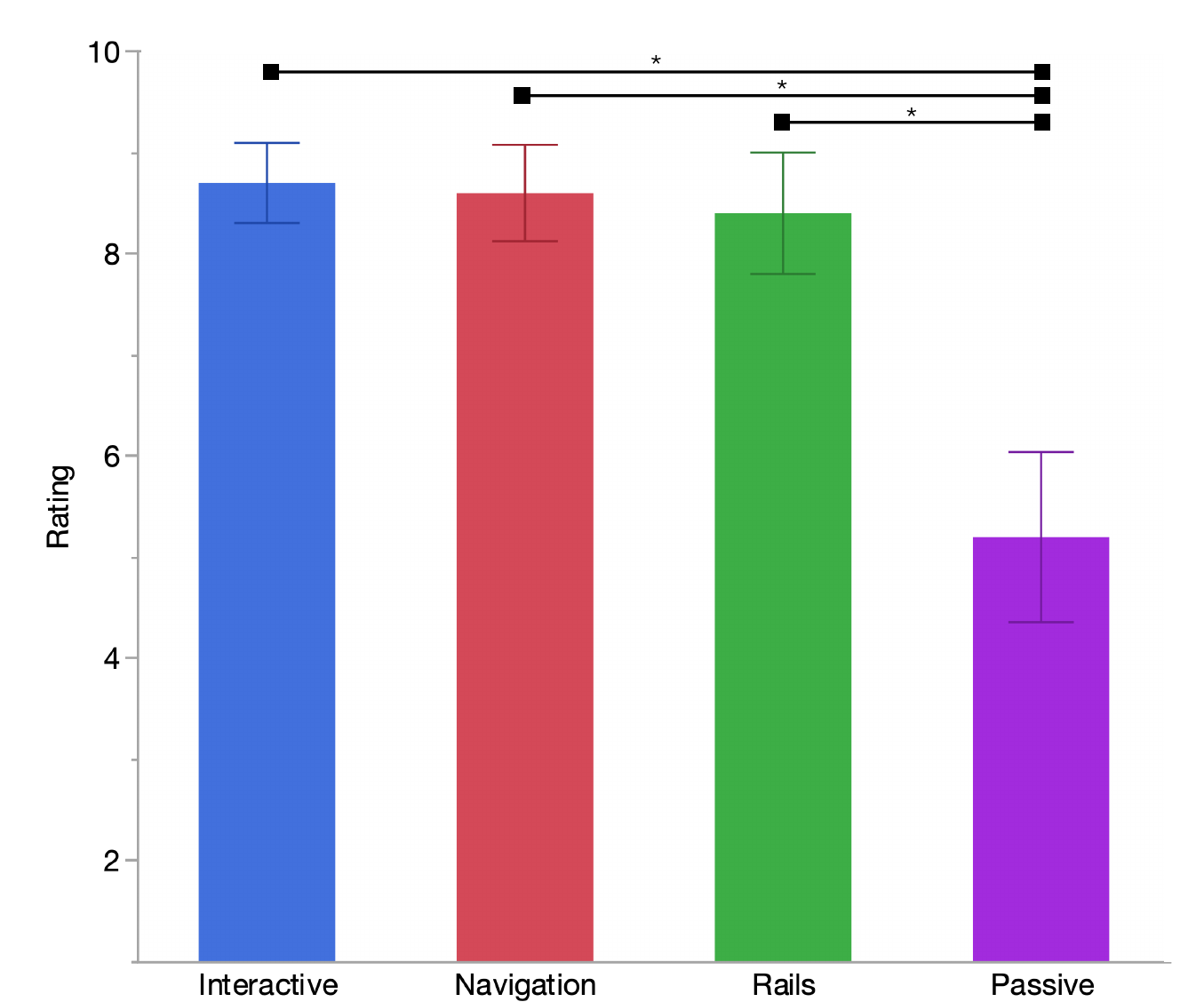}
        \caption{ }
        \label{fig:ratings_plot}
    \end{subfigure}
    \setlength{\belowcaptionskip}{-15pt}
    \vspace*{-15pt}
    \caption{Bar charts of questionnaire scores. Significantly different pairs are marked with * when $\emph{p} \le 0.05$ and ** when $\emph{p} \le 0.01$. Bar chart whiskers are the $\pm S.E.$ spread of the data. (a): Mean SSQ scores from pre- and post-study. (b): Mean UES scores. (c): Mean NASA-TLX scores. (d): Mean condition ratings.}
\end{figure*}

Using the UES-SF data, Kruskal-Wallis results found statistically significant score differences for \textit{Aesthetic Appeal} ($H(3)=21.61$, $\emph{p}<0.01$), \textit{Perceived Usability} ($H(3)=11.90$, $\emph{p}<0.01$), and \textit{Reward} ($H(3)=21.61$, $\emph{p}<0.01$) dimensions. 
From post-hoc pairwise comparisons for \textit{Aesthetic Appeal}, we found the \textsc{Passive} interaction condition to be significantly less appealing than the \textsc{Interactive} ($\emph{p}<0.01$), \textsc{Navigation} ($\emph{p}<0.01$), and \textsc{Rails} ($\emph{p}<0.01$) conditions.
For \textit{Perceived Usability}, we found the \textsc{Passive} interaction condition to be significantly less usable than the \textsc{Interactive} ($\emph{p}=0.03$), \textsc{Navigation} ($\emph{p}=0.04$), and \textsc{Rails} ($\emph{p}=0.02$) conditions.
Finally, for the \textit{Reward}, we found the \textsc{Passive} interaction condition to be significantly less rewarding than the \textsc{Interactive} ($\emph{p}<0.01$), \textsc{Navigation} ($\emph{p}<0.01$), and \textsc{Rails} ($\emph{p}<0.01$) conditions.
These significant pairs and mean UES-SF scale responses are shown in Fig. \ref{fig:ues_plot}.

\subsection{Workload}
Kruskal-Wallis tests across the NASA-TLX dimensions reported significant differences between interaction conditions for the \textit{Mental Demand} ($H(3)=8.84$, $\emph{p}=0.03$) and \textit{Frustration} ($H(3)=9.49$, $\emph{p}=0.02$) dimensions.
After conducting post-hoc pairwise comparisons using Mann-Whitney U tests with Holm-Bonferroni corrections, no interaction condition pairs were found to be significantly different for the \textit{Mental Demand} dimension.
For the \textit{Frustration} dimension, the \textsc{Passive} condition was found to be significantly more frustrating than the \textsc{Interactive} condition ($\emph{p}=0.05$).
The mean scores for the NASA-TLX are shown in Fig. \ref{fig:tlx_plot}.

\subsection{Ratings}
Kruskal-Wallis tests on the reported ratings of interaction conditions indicated significant differences between them ($H(3)=13.13$, $\emph{p}<0.01$).
From post-hoc pairwise comparisons, we found that the \textsc{Passive} interaction condition was rated significantly lower than the \textsc{Interactive} ($\emph{p}=0.02$), \textsc{Navigation} ($\emph{p}=0.02$), and \textsc{Rails} ($\emph{p}=0.03$) conditions.
These significant pairs and mean condition ratings are shown in Fig. \ref{fig:ratings_plot}.

\begin{figure*}[tb]
    \centering
    \begin{subfigure}[b]{\columnwidth}
        \centering
        \includegraphics[width=\columnwidth]{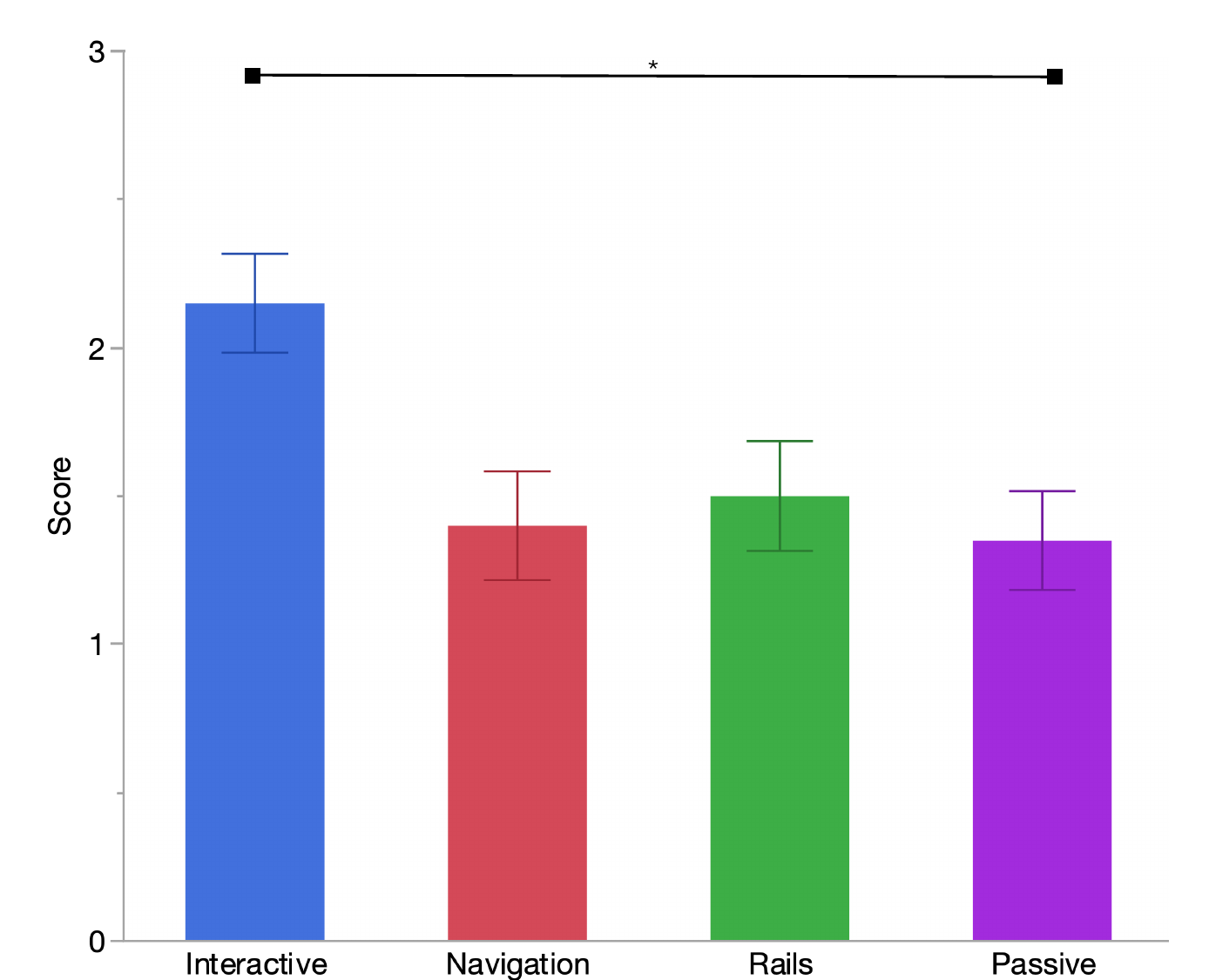}
        \caption{ }
        \label{fig:verbal_plot}
    \end{subfigure}
    \hspace{2.1em}
    \begin{subfigure}[b]{\columnwidth}
        \centering
        \includegraphics[width=\columnwidth]{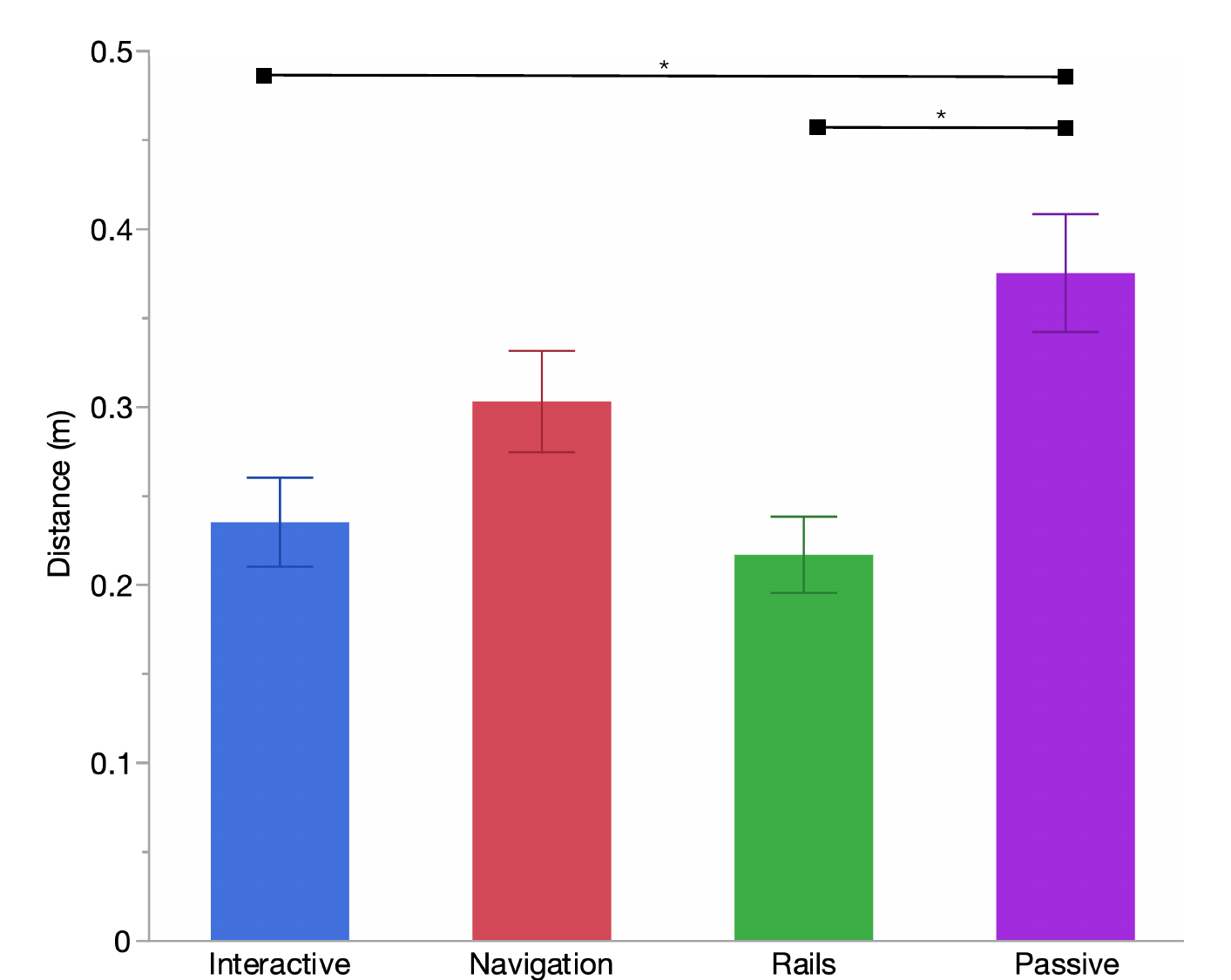}
        \caption{ }
        \label{fig:spatial_plot}
    \end{subfigure}
    \setlength{\belowcaptionskip}{-15pt}
    \vspace*{-15pt}
    \caption{Bar charts of collected test responses for verbal and spatial questions from Table \ref{tbl:tests}.  (a): Verbal scores. (b): Distances of participant-reported defect location from the actual defect location.}
\end{figure*}

\subsection{Auditory Recall}
We graded participant scores based on the transcribed responses to the verbal test questions posed (see Table \ref{tbl:tests}).
Responses were marked as either correct or incorrect based on a common grading rubric, with no partial credit being awarded.
A Shapiro-Wilk test was used to check the normality of the test scores for each interaction condition, revealing that the data was not normally distributed.
We then applied an Aligned Rank Transform (ART) before using a mixed Analysis of Variance (ANOVA), as the \textit{retention time} (i.e., short and medium-term) was a within-subjects factor and the interactive condition a between-subjects factor \cite{Wobbrock_ArtAnova2011}.
The ART ANOVA indicated a significant effect of the interactive condition ($F(3, 36) = 3.16$, $ \emph{p} = 0.04$) and no effect of retention time ($F(1, 36) = 1.00$, $ \emph{p} = 0.32$) or the interaction between the interactive condition and retention time ($F(3, 36) = 0.21$, $ \emph{p} = 0.89$).
Post hoc pairwise comparisons were then conducted for the interactive condition factor with the ART-C procedure, and corrected with Tukey's Honest Significance Difference (HSD) \cite{Elkin_ArtC2021}.
This showed that the difference in audio recall scores between the \textsc{Interactive} and \textsc{Passive} conditions were statistically significant ($p = 0.04$).
No other pairwise comparisons were significantly different.
The mean scores across interactive conditions and the significant pairing are shown in Fig. \ref{fig:verbal_plot}.

\subsection{Spatial Recall}
A Shapiro-Wilk test was first used to check the normality of reported defect distances for each interaction condition, revealing that the data was not normally distributed.
We then applied an ART before using a mixed ANOVA for the within-subjects factor of retention time and the between-subjects factor of interactive condition.
The ART ANOVA indicated a significant effect of the interactive condition ($F(3, 36) = 4.18$, $ \emph{p} = 0.01$) and no effect of retention time ($F(1, 36) = 1.30$, $ \emph{p} = 0.26$) or the interaction between the interactive condition and retention time ($F(3, 36) = 1.72$, $ \emph{p} = 0.18$).
Post hoc pairwise comparisons were then conducted for the interactive condition factor with the ART-C procedure, and corrected with Tukey's HSD.
This showed that the difference in spatial recall distances between the \textsc{Interactive} and \textsc{Passive} conditions were statistically significant ($p = 0.04$), as well as the \textsc{Rails} and \textsc{Passive} conditions ($p = 0.02$).
No other pairwise comparisons were significantly different.
The mean reported distances and significance pairings are shown in Fig. \ref{fig:spatial_plot}.

\section{Discussion}

\subsection{Recall}
To assess the participant's recollection of information covered during the guided tour, we created a set of test questions (i.e., Table \ref{tbl:tests}) regarding the locations of defects within the 3D object and dialogue spoken by the guide throughout the tour.
These tests were administered immediately following the tour and after a buffer activity, as described in Sec. \ref{sec:procedure}.
This was intended to capture their retention of the tour in the short term and following a verbal and visuospatial buffer \cite{Jonides_MindMemory2008}.
We used the test responses and open-interview transcripts to answer \textbf{RQ1} and test the corresponding hypotheses: \textit{H1.1} and \textit{H1.2}.

We hypothesized that participants who experienced the tour with the \textsc{Interactive} condition would have a greater recollection of verbal information than the other conditions (\textit{H1.1}).
Our results partially support \textit{H1.1}, as those who participated in the \textsc{Interactive} tour achieved significantly higher scores compared to those in the \textsc{Passive} condition.
However, despite \textsc{Interactive} achieving the highest average score, it was not found to be significantly different than \textsc{Navigation} and \textsc{Rails}.
Although all participants listened to the same tour narrative from the guide, this result suggests that interaction during the embodied inspection proved crucial in retaining the auditory information presented.
We believe this follows the view in sensorimotor enactivism that one's perception relies on the active exploration of an environment, consisting of movements and sensory states in that environment, where the participant's auditory perception was improved by the additional playback actions \cite{Shapiro_EmbodiedCog2024}.
The \textit{Controller Point} action was described by several participants in the \textsc{Interactive} condition as helping them direct their attention to and remember areas of interest.
\textit{P5} mentioned ``the guide telling me to look at the point... I feel like I still might have missed it unless I actually went into that position and put my hand exactly where it was pointing'', \textit{P17} stated that they ``liked when the guide made me point... it's clicking to the point and being like `Yeah, I see it'... it's like we're in the same spot'', and \textit{P37} described that when ``the tour guide sees something you don't quite see yet, they point at it... you kind of get that `aha' moment of like, ohh, that's what we're looking at.'' 
From the other tour conditions, participants described their desire to further interact with the environment influencing their retention of information.
\textit{P6} in the \textsc{Navigation} condition mentioned ``if I was able to interact more... I think I would have been able to retain the information a little better'', \textit{P31} of the \textsc{Rails} condition described preferring ``to have some time for myself to navigate through the object after [the guide] explained... I could test and review the object for myself or see that information'', and \textit{P12} in the \textsc{Passive} condition stated ``I prefer being more hands-on with things and I wish I had control as far as like when I could proceed and I almost wish there was a pause function so you could, like, stop and look instead of here's this thing, here's that thing.''

We additionally hypothesized that participants who experienced the tour with the \textsc{Interactive} condition would have a greater recollection of spatial information than all other conditions (\textit{H1.2}).
Our results partially support \textit{H1.2}, as those who participated in the \textsc{Interactive} tour achieved significantly better estimates of the approximate locations of defects compared to those in the \textsc{Passive} condition.
However, the same statistical significance was found for the \textsc{Rails} condition when compared to the \textsc{Passive} condition.
This is similar to a comparison study examining viewpoint control, where the participants who had independent control over their viewpoint achieved higher test scores than those with a fixed viewpoint \cite{Lovreglio_Training2021}.
This implies a greater influence of viewpoint in visuospatial recall, rather than our anticipated grounding that the participant benefits from body movements required by our interactions to increase environment understanding \cite{Shapiro_EmbodiedCog2024}.
A missing pairing of interest was the significance between the \textsc{Navigation} and \textsc{Passive} conditions, where we would similarly expect the independent control of viewpoint resulting in better recollection of the approximate positions of defects.
We attribute this to our guided tour system, which allowed participants to perform the \textit{Navigation Down} and \textit{Navigation Up} at any point during the \textsc{Interactive} and \textsc{Navigation} conditions, resulting in many participants preemptively navigating across scales, requiring them to navigate back to the guide to resume playback.
A few participants described this in the post-study interview, such as \textit{P21} from the \textsc{Interactive} condition stating that their ``only issue would be going too fast. Hearing like `let's zoom in' and then doing it before I heard the pause sound effect'' and \textit{P42} in the \textsc{Navigation} condition describing ``I got too curious at some points and when the guy was guiding me to a different point of interest for the struts I was like `I really want to go see it' and I would. I did not wait and went rogue at some points.''

Overall, the tours using the \textsc{Interactive} condition had a greater auditory recall and better spatial recall than the \textsc{Passive} condition.
Although the \textsc{Interactive} condition did not show significantly better recall compared to the other independent viewpoint conditions, it produced the highest performance, which could potentially be enhanced with adjustments to the playback system.
The evidence, therefore, suggests that additional interactivity beyond simple viewpoint control can help increase observers' recall of guided tour content.

\subsection{User Experience}
To assess the participants' experience from following the guided tour, we leveraged existing research questionnaires including the SSQ, UES-SF, and NASA-TLX, as well as post-study ratings from 1-10 if participants would participate in another tour in the same format they experienced.
These questionnaires were completed by the participant throughout the study as documented in the procedure.
We used the responses and open-interview transcripts to answer \textbf{RQ2} and test the affiliated hypotheses: \textit{H2.1}, \textit{H2.2}, and \textit{H2.3}.

We hypothesized that the \textsc{Interactive} and \textsc{Navigation} conditions would have higher reported engagement than the \textsc{Rails} and \textsc{Passive} conditions (\textit{H2.1}).
Based on the UES-SF response data, our results partially support \textit{H2.1} with an interesting exception.
The \textsc{Rails} condition, although not having controller-based interaction, was reported as being approximately as engaging as the \textsc{Interactive} and \textsc{Navigation} conditions, with significance existing from higher reported scores across \textit{Aesthetic Appeal}, \textit{Perceived Usability}, and \textit{Reward} dimensions of the UES-SF than the \textsc{Passive} condition.
Reviewing open-interview transcripts from participants who completed the \textsc{Passive} condition, the reasons for these lower scores could be found.
For instance, lack of engagement using VR was an issue mentioned by several participants, such as \textit{P12} who stated ``it was almost kind of like you were doing it in VR and then you were kind of being drug around'', \textit{P36} describing ``for me it's about control. I didn't have that during the video... didn't have my own movement and head movements'', and \textit{P44} saying ``watching a video in VR kind of defeats the purpose... since it's just a video, I'm not interacting with anything.''

We also hypothesized that the \textsc{Passive} condition would be reported as having a higher task workload than all other conditions (\textit{H2.2}).
From the NASA-TLX response data, our results do not support \textit{H2.2}.
While the \textsc{Passive} condition did have worse task load scores across all dimensions of the NASA-TLX, only the \textit{Frustration} dimension had a significant pairing with the \textsc{Interaction} condition.
Still, participants explicitly mentioned issues with task load when discussing the \textsc{Passive} condition in the post-study interview.
Regarding the \textit{Temporal} load, \textit{P40} described feeling the tour ``was fast... a lot of information in a very short time'' and \textit{P28} stated that the tour was ``a bit too demanding and I think it was a bit fast-paced, so kind of keeping track was a bit difficult.''
Additional dimensions mentioned were the \textit{Mental} load, where \textit{P36} stated that ``to memorize someone else's movement and what they were saying and what I was seeing... that created a lot of mental load'', and the \textit{Frustration}, with \textit{P12} remarking that the tour was simply ``really hard to follow.''

Finally, we hypothesized that the \textsc{Rails} and \textsc{Passive} conditions would have higher reported simulator sickness scores than the \textsc{Interactive} and \textsc{Navigation} conditions (\textit{H2.3}).
From the SSQ response data, our results do not support \textit{H2.3}.
Although the \textsc{Passive} condition did have significant increases in simulator sickness scores between pre- and post-study measurements across all dimensions, the \textsc{Rails} condition did not have any significant increases.
Furthermore, the \textsc{Interactive} condition reported having significant increases in simulator sickness for the \textit{Oculomotor} and \textit{Total} dimensions.
We attribute this to \textit{WYSIWIS}, where participants reported discomfort when performing the playback interaction.
\textit{P13} stated that they ``didn't love [WYSIWIS]... having the transparent yellow headset and having to look... my view was being stuck on the headset'', \textit{P17} described that the ``[WYSIWIS] blocked my view and it was weird'', and \textit{P25} said ``the headset one... is a little disorienting.''

Overall, the tour using the \textsc{Passive} condition was found to have a worse user experience across multiple engagement, task workload, and simulator sickness measures.
Along with their final ratings, participants found the \textsc{Passive} condition to be limited in its usability and subsequent engagement.
Conversely, the other conditions were comparably rated across these measures, making them all viable means to conduct tours from the user's perspective.
We attribute this difference to the distinct design of the \textsc{Passive} condition, in which participants did not control the tour's viewpoint, experiencing it instead from a third-person perspective.
This implies a greater influence of viewpoint on the user's experience as opposed to strictly the extent of interaction during the guided tour we expected.

\subsection{Limitations \& Future Work}
Our work had several limitations. 
First, the subject matter of 3D printing was not uniformly understood or of interest to the participants across conditions.
This could have affected the recall and experience of the participant.
Future work could consist of a more generalized topic outside of advanced manufacturing to be exciting and more universally understood, such as sports coach-player mentoring  \cite{Lin_Basketball2021}, medicine including anatomy and immunology \cite{Zhao_effectiveness2020, Zhang_InteractivityStory2019}, or general education topics, such as history and cultural heritage \cite{Roussou_LearningByDoing2004}.
Secondly, the sample size of our study was relatively small due to the between-subjects design.
A larger sample size may have resulted in us finding significance in cases where values were trending towards a significant difference, such as the NASA-TLX dimensions.
Future studies should recruit more participants to thoroughly analyze the recall outcomes and user experience.
Alternatively, the same participant quantity could be used by omitting a condition for comparison, providing more data to analyze between the remaining conditions.
Additional tours within the VE for unique inspection objects could also provide a means to design a within-subjects experiment.
Thirdly, the asynchronous collaboration was focused on a confederate recording performed by the investigator and viewed by the observing participant.
A future study could focus on having a preceding participant perform a recording for the following participant to observe, such that insights regarding the recording production process can be considered.
Lastly, the 3D printed object used for inspection was homogenous due to the unit cell models it was comprised of.
Future studies could consider using an object with distinct differences in its geometry or markers to provide reference points for determining relative locations of points of interest covered during a tour.

\section{Conclusion}
In this work, we document our design of a guided tour system to facilitate asynchronous collaboration for AM inspections.
We describe the use of guide mimicry by the observing user as a means to progress the playback of a guided tour.
Our results from a user study investigating the extent of using these mimicking actions suggest that interactivity positively influences auditory and spatial recall and that the extent of viewpoint control helps improve the user's experience when following a guided tour.
For future work, we plan to investigate the use of guided tours in more generalized collaborative topics.

\acknowledgments{
This work was supported by the US DOE LLNL-LDRD 23-SI-003 and was performed under the auspices of the U.S. Department of Energy by Lawrence Livermore National Laboratory under Contract DE-AC52-07NA27344 (LLNL-CONF-869200).
}

\bibliographystyle{abbrv-doi}

\bibliography{template}
\end{document}